\documentclass{aastex}
\usepackage{emulateapj5} 
\usepackage{apjfonts}

\shorttitle{AO Spectroscopy of IRAS~22272 in 3~$\mu$m} 
\shortauthors{Goto et al.}
\begin{document}



\title{Spatially Resolved 3~\micron~Spectroscopy of IRAS~22272+5435:
\\ Formation and Evolution of Aliphatic Hydrocarbon Dust \\ in 
Proto-Planetary Nebula\altaffilmark{1}}

\altaffiltext{1}{Based on data collected at Subaru Telescope,
which is operated by the National Astronomical Observatory of Japan.}

 \author{Miwa Goto, \altaffilmark{2,3} 
W. Gaessler,\altaffilmark{2,4}
Yutaka Hayano,\altaffilmark{5} Masanori Iye,\altaffilmark{5} 
Yukiko Kamata,\altaffilmark{5} Tomio Kanzawa,\altaffilmark{2} 
Naoto Kobayashi,\altaffilmark{2} Yosuke Minowa,\altaffilmark{5}
D. J. Saint-Jacques,\altaffilmark{6} Hideki Takami,\altaffilmark{2}
Naruhisa Takato,\altaffilmark{2} Hiroshi Terada,\altaffilmark{2}}

\email{mgoto@duke.ifa.hawaii.edu}

\altaffiltext{2}{Subaru Telescope, 650 North A`ohoku Place, Hilo,
HI 96720.}

\altaffiltext{3}{Visiting astronomer at the Institute for
Astronomy, University of Hawaii.}


\altaffiltext{4}{Max-Planck-Institut f\"ur Astronomie, K\"onigstuhl 17,
Heidelberg D-69117, Germany.}

\altaffiltext{5}{National Astronomical Observatory of Japan, Mitaka,
Tokyo 181-8588, Japan.}

\altaffiltext{6}{D\'epartment de physique, Universit\'e de
Montr\'eal, Montr\'eal (Qu\'ebec) H3C 3J7, Canada.}






\begin{abstract}
We present medium-resolution 3~$\mu$m spectroscopy of the carbon-rich
proto-planetary nebula IRAS~22272+5435. Spectroscopy with the Subaru
Telescope adaptive optics system revealed a spatial variation of
hydrocarbon molecules and dust surrounding the star. The
ro-vibrational bands of acetylene (C$_2$H$_2$) and hydrogen cyanide
(HCN) at 3.0~$\mu$m are evident in the central star spectra. The
molecules are concentrated in the compact region near the center. The
3.3 and 3.4~$\mu$m emission of aromatic and aliphatic hydrocarbons is
detected at 600--1300~AU from the central star. The separation of
spatial distribution between gas and dust suggests that the small
hydrocarbon molecules are indeed the source of solid material, and
that the gas leftover from the grain formation is being observed near
the central star. The intensity of aliphatic hydrocarbon emission
relative to the aromatic hydrocarbon emission decreases with distance
from the central star. The spectral variation is well matched to that
of a laboratory analog thermally annealed with different
temperatures. We suggest that either the thermal process after the
formation of a grain or the variation in the temperature in the
dust-forming region over time determines the chemical composition of
the hydrocarbon dust around the proto-planetary nebula.

\end{abstract}

\section{Introduction} 

IRAS~22272+5435 (= HD~285858, SAO~34504) is an extremely carbon-rich
proto-planetary nebula (PPN) with peculiar infrared spectral features
\citep{kwo89,geb92}. A PPN is a transition object that has left the
asymptotic giant branch (AGB) for a planetary nebula (PN), but it is
not yet hot enough to ionize the surrounding AGB ejecta. The dust in a
PPN is becoming optically thin, but has not been altered in the hard
UV field of a PN. A carbon-rich PPN provides the best opportunity to
study unprocessed solid carbon material immediately after the
formation.

A classification of carbon dust in the interstellar medium (ISM) has
been proposed by \citet{geb97} and \citet{tok97} based on the 3~$\mu$m
infrared emission features (IEF). The class A sources are
characterized by the intense and isolated aromatic features at
3.3~$\mu$m with an occasional association of weak and sharp peaks in
the 3.4--3.5~$\mu$m region. They are the most common IEF observed
toward young stellar objects (YSO), \ion{H}{2} regions, PN, reflection
nebulae, diffuse interstellar medium, and extragalactic sources
\citep{tok91,slo97,jou90,sel83,tan96,moo96}. The class B sources are
characterized by broad and prominent aliphatic features at 3.4~$\mu$m
accompanied by the 3.3~$\mu$m aromatic features. The class B IEF has
been found exclusively toward PPNs with extremely carbon-rich
chemistry typified by IRAS~22272+5435 \citep{geb90,geb92}. The
simultaneous presence of aromatic and aliphatic features in class B
sources makes a good laboratory for the study of the transition of the
chemical composition of hydrocarbons.

Another advantage in observing a PPN is that the history of its dust
formation is locked in the spatial distribution of the material. Since
PPNs are less developed than PNs, most of them are spatially
unresolved. However, the dust distribution around IRAS~22272+5435 has
been well studied by subarcsecond mid-infrared imaging
\citep{mei97,day98,uet01}, optical imaging with the {\it Hubble Space
Telescope}/Wide Field Planetary Camera 2 ({\it HST}/WFPC2)
\citep{uet00}, and near-infrared polarimetric imaging
\citep{gle01}. Ueta et al. (2000) classified IRAS~22272+5435 as a
member of PPNs that are optically thin. It therefore ensures an
unobscured view of the central part of the nebula.

Although observing at 3~$\mu$m is essential to assess the chemical
composition of the carbon dust, a drawback is that the emitting region
is physically more compact than at longer wavelengths. In addition,
the contribution of the direct and scattered light from the central
star is not negligible in the 3~$\mu$m region. Both to resolve the
dust emission region and to achieve the high dynamic range required to
distinguish the faint dust emission from the central star, an adaptive
optics (AO) system must be integral part of the instruments used to
make the spectroscopic observations.  Our goal for this work is to use
the high spatial resolution spectroscopy afforded by AO to better
understand how hydrocarbon dust forms around a carbon-rich evolved
star, and how it is processed before dissipating into the local ISM.
 
\section{Observation} 

The spectroscopic observation was made on UT 2001 July 13 using the
Infrared Camera and Spectrograph (IRCS; Tokunaga et al. 1998;
Kobayashi et al. 2000) at the 8.2~m Subaru Telescope in conjunction
with the AO system \citep{tak98,gae01}. The Subaru AO system is a
36-element curvature system installed at the front end of the
telescope Cassegrain port. A medium-resolution grism was used with a
0\farcs30 slit in the 58 mas camera section to provide spectra from
2.84 to 4.18~$\mu$m with a resolving power of 600--800. The visible
central star
was used as the wavefront reference source for the AO system.  The
position angle of the slit was 56$^\circ$ along the elongation in the
mid-infrared image that corresponds to an equatorial density
enhancement of the nebula \citep{uet01}. The spectra were recorded by
nodding the tip-tilt mirror inside the AO system by 2\arcsec~along the
slit to subtract the sky emission and dark current images. The total
on-source integration time was 720 s. A nearby F8V star HR~8472 was
observed as a spectroscopic standard at similar airmass. The
spectroscopic flat field was obtained at the end of the night with a
halogen lamp.

\section{Data Reduction and Results}

We obtained the one-dimensional spectra of IRAS~22272+5435 using the
IRAF\footnote{IRAF is distributed by the National Optical Astronomy
Observatories, which are operated by the Association of Universities
for Research in Astronomy, Inc., under a cooperative agreement with
the National Science Foundation.} aperture extraction package. The
aperture width was about equal to the FWHM of the spatial profile
(4-pixel or 0\farcs23). We set the extraction apertures at the central
star and another 14 locations along the slit (Figure~\ref{prof}). We
found the spectra recorded at the nodding positions A and B to be
slightly different. The most likely reason for this is that the mirror
nodding axis was not precisely straight to the slit, and we observed
slightly different parts of the nebula. The data obtained at positions
A and B were reduced separately. The wavelength calibration was
performed by fitting over 60 telluric absorption lines in the 3~$\mu$m
region. The flux calibration relative to the standard star was made
assuming the intrinsic spectrum of a F8V star is represented by a
Planck function of $T_{\rm eff} = $ 6100~K \citep{tok00}. The hydrogen
recombination lines in the standard star spectrum were fit with a
Gaussian function and subtracted before dividing. The details of the
reduction can be found elsewhere \citep{got02}.

The result is shown in Figure~\ref{spec1} after being normalized to
the continuum at 3.5~$\mu$m. No hydrocarbon emission feature is seen
at the position of the central star. Instead we observe absorption
bands with the hydrocarbon emission bands becoming evident only at
0\farcs35 from the central star. The integrated intensity of the
hydrocarbon emission bands is consistent with that of
5\arcsec~aperture spectroscopy formerly reported by \citet{geb92}. The
absence of hydrogen recombination lines at 3.74~$\mu$m (Pf$\gamma$)
and 4.05~$\mu$m (Br$\alpha$) ensures that no \ion{H}{2} region has
developed at the center of the nebula. The spectrum continuum is
approximately represented by a power law. The continuum flux around
the central star is most likely the light scattered by large dust
grains in the nebula superposed on the halo component of the
point-spread function of the central star. The continuum subtracted
spectra are shown in Figure~\ref{spec2}.

\section{Discussion}
\subsection{Molecular Absorption Features near the Center}

A wealth of absorption features is found in the spectra of
IRAS~22272+5435 close to the central star. The carbon atoms consisting
of C-H bonds have one of $sp$, $sp^2$, or $sp^3$ hybridized orbitals,
in which triple, double, and single bonds connect the carbon with
adjacent atoms. The stretching vibration mode of different C-H bonds
shows up as various band structures in the 3~$\mu$m region. The
individual ro-vibrational transition lines are not resolved with the
present spectral resolution, and higher resolution spectroscopy will
be required to make a solid identification of the absorption
lines. The only exception is the absorption bands at 3.0~$\mu$m, which
are identified with the combination of $P$ and $R$ branches of $sp$
C-H bonds in C$_2$H$_2$ and HCN. Figure~\ref{mol2} shows the same
molecular features in CRL~618 \citep{chi98} and those computed with
the HITRAN database \citep{rot98}.

The molecular absorption features are evident close to the central
star, but they disappear abruptly at 0\farcs47 or 750~AU ($d$ =
1.6~kpc is assumed based on a private communication with Nakashima
2002). The broad emission bands at 3.3 and 3.4~$\mu$m become strong in
turn at 600~AU (Fig.~\ref{spec2}). This is the first time that both
the hydrocarbon gas and dust are observed in a circumstellar
environment in clearly separated locations. It indicates the
hydrocarbon molecules are indeed the source of the solid dust
material, and that the molecular gas that failed to be involved with
the grain formation is being observed near the central star.

\subsection{Hydrocarbon Dust Emission at 3.3--3.4~$\mu$m}

The most common IEF in the ISM is the class A type emission
ubiquitously observed where the carbon dust is exposed to a hard UV
field. The emission carrier has been attributed to aromatic species
\citep{dul81} such as neutral and ionized polycyclic aromatic
hydrocarbons (PAH; L\'{e}ger \& Puget 1984; Allamandola, Tielens, \&
Barker 1985, 1989 and references therein). The 3.3~$\mu$m feature
matches the ro-vibrational band of the stretching vibration of
aromatic $sp^2$ C-H bonds. While the class A IEF is observed with hot
excitation sources with $T_{\rm eff} > $ 20000~K, the central stars of
class B sources are typically F--G type supergiants whose effective
temperature is $T_{\rm eff}$ = 4500--7000~K. The radiation environment
of class B sources is much more benign with less UV photons
available. The aromatic molecules absorbs UV-visible photons with a
long wavelength cut-off. The larger molecules have a longer cut-off
wavelength \citep{des90,sch93}. The soft radiation field yet strong
emission of class B features implies larger hydrocarbons than those of
class A sources are responsible.

What distinguishes class B spectra most clearly from class A IEF is
broad and intense emission at 3.4~$\mu$m. It was once proposed that a
hot band of anharmonic oscillation of aromatic C-H bonds is
responsible \citep{bar87}. However, the weakness or absence of
overtone bands at the 1.6--1.8~$\mu$m region makes this hypothesis
unlikely \citep{mag92,sie93,geb94}. Instead \citet{sch93} proposed
that the aliphatic C-H bonds of superhydrogenated benzene rings
account for the 3.4~$\mu$m emission. Hydrogenated PAH (H$_n$-PAH) is
rich in $sp^3$ C-H bonds where the double bonds in the benzene rings
are partially saturated by extra attachment of hydrogens.
\citet{ber96} have shown that some H$_n$-PAH have similar absorption
peaks that agree well with 3.4--3.5~$\mu$m features of a class A
source Orion bar. \citet{job96} argue that the aliphatic C-H bonds, in
particular -CH$_2$ in long chains or cyclic aliphatic hydrocarbons,
best match the strong 3.42~$\mu$m features of class B source
IRAS~05341+0852.

In the mid-infrared region, IRAS~22272+5435 shows two broad emission
features near 8~$\mu$m and 12~$\mu$m \citep{bus93}, which contrasts
with typical class A IEF \citep{slo99}. \citet{all99} tried several
laboratory compounds of PAH to reproduce IEF of class A and class B
sources. They suggested that there is a systematic difference in the
character of aromatic species required to fit the IEF in the two
different environments. The class A IEF is most closely reproduced by
compounds of cations of stable PAH species, while neutral and less
stable PAH species are a better match to the broad features of class B
sources.

Specific chemical materials have been proposed for the class B IEF
carriers by many authors. Those substances are in a more solid form,
and laboratory spectroscopy indicates they are rich in $sp^3$
hybridized bonds. These are the more common characteristics of
amorphous carbons, totally fit in the above modifications required for
the conventional aromatic species, namely, larger, more neutral, and
more aliphatic materials for class B IEF. The possible carriers
presented so far are all successful in reproducing the spectral shape
from 3.3 to 3.4~$\mu$m. These include a mixture of PAH molecules fully
reprocessed by energetic hydrogen plasma \citep{bee97,arn00}, carbon
nanoparticles produced by laser pyrolysis of C$_2$H$_4$ and C$_4$H$_6$
\citep{her98} or C$_2$H$_2$ \citep{sch99}, hydrogenated amorphous
carbon prepared by eximer laser ablation of graphite in a hydrogen
rich atmosphere \citep{sco97, gri00}, and quenched carbonaceous
composite produced by the plasma deposition technique with CH$_4$
\citep{sak90,got00}. Semianthracite coal grains \citep{gui96} are also
successful in reproducing the aliphatic band at 3.4~$\mu$m and
substructures in the mid-infrared feature and overall continuum
shape. Since there is no unique laboratory analog, we refer the
carrier of the class B emission features at 3.3--3.4~$\mu$m as
``hydrocarbon dust'' in the following text. The hydrocarbon dust is
rich in both aromatic and aliphatic C-H bonds.

\subsection{Variations of Aliphatic and Aromatic Abundance}

Figure~\ref{labo} shows the emission feature at 600 to 1300 AU from
the central star at the northeast sector of the nebula. The relative
intensity of the aliphatic feature at 3.4~$\mu$m to the aromatic
feature at 3.3~$\mu$m decreases with the distance from the central
star. We discuss two processes that could modify the relative
abundance of the aliphatics to the aromatics.

First, we examine the formation of the aliphatic C-H bonds on the
surface of carbon particles. \citet{men99} demonstrated that $sp^3$
hybridized C-H bonds are newly formed by exposing pure carbon
particles to an atomic hydrogen atmosphere. The activated aliphatic
band successfully reproduces the 3.4~$\mu$m absorption feature
observed toward the Galactic center. They discuss that the carrier
responsible for the 3.4~$\mu$m absorption ubiquitously observed toward
highly reddened objects does not originate in cold dark clouds as has
been proposed by \citet{gre95}, but in the diffuse ISM where the
destruction of C-H bonds by UV photolysis is equilibrated by
rehydrogenation with ample atomic hydrogen \citep{men01,men02}.
\citet{tie94} proposed a similar mechanism for the formation of
aliphatic C-H bonds by hydrogenation of a graphite surface amorphized
by the ion bombardment in the interstellar shocks.  The process
supposedly plays an important role in the production of the 3.4~$\mu$m
absorption feature in the Galactic center sources
\citep{sch99}. \citet{chi98} proposed that the shocks caused by the
fast wind at the last stage of the post-AGB evolution is relevant to
the formation of the 3.4~$\mu$m absorption carriers in CRL~618.

However, the freshly formed aliphatic bands do not match the {\it
emission} feature we observed. In Figure~\ref{men} we compare the
emission feature of IRAS~22272+5435 with the absorption features of
carbon particles from \citet{men99}, CRL~618 \citep{chi98} and the
Galactic center \citep{chi00}, as well as the laboratory analog of
\citet{got00}. Indeed, the hydrogenated nano-size carbon particles of
\citet{men99} reproduce the observed {\it absorption} band well; the
discrepancy with IRAS~22272+5435 is obvious, indicating different
carriers. Considering CRL~618 is a PPN more evolved than
IRAS~22272+5435 with a compact \ion{H}{2} region and a hot B0 star at
the center of the nebula, the availability of UV might be critical to
producing the particular chemical composition of the absorbing aliphatic
hydrocarbons.

The thermal annealing of aliphatic material better fits the observed
spectral variation. Figure~\ref{labo} shows the 3~$\mu$m spectra of a
laboratory analog of hydrocarbon dust subjected to thermal annealing
at different temperatures \citep{got00}. The model carbon dust
produced by the plasma vapor deposition method of CH$_4$ has rich
$sp^3$ hybridized C-H bonds when it is deposited. As the annealing
temperature becomes hotter in the postprocessing, the substance is
dehydrogenated and graphitized. It gradually acquires strong continuum
absorption, less prominent spectral features, and weak aliphatic
features relative to the aromatic one. The thermal transformation of
the aliphatic material is well studied in laboratory experiments with
amorphous carbon films \citep{rob91,boun95}. Similar spectral
alterations have been also reported with hydrogenated amorphous
carbons \citep{sco96,sco97,gri00,men96}. The resemblance between the
laboratory and the observed spectra indicates a similar process takes
place in IRAS~22272+5435.


\subsection{Thermal Process}
\subsubsection{Heat Source}
For the thermal annealing of aliphatic material to take place, we need
an adequate heat source for the raw material. The chemical composition
of the thermally processed material is sensitive to the annealing
temperature. The aliphatic to aromatic ratio observed in
IRAS~22272+5435 indicates that the hydrocarbon dust has undergone
heating to about 760~K (Figure~\ref{labo}).

In the investigation of heating mechanisms, we first assumed the
thermal annealing is currently in progress at $\sim$1000~AU away from
the star, where the hydrocarbon emission is being observed. The
equilibrium temperature at 1000~AU is too cool for annealing to
happen. \citet{uet01} solved the two-dimensional radiative transfer
model to get $T_{\rm dust}$ = 200~K at 800~AU from the star, which is
again too cool for the thermal annealing. Shock heating could be an
alternative heat source; however, no sign of shock-induced molecular
hydrogen emission has been observed in the 2~$\mu$m spectrum
\citep{hri94}.

The heating mechanism is not necessarily a thermal equilibrium
process. Thermal fluctuation or stochastic heating occurs when the
energy of a single incident photon exceeds the total heat capacity of
a grain. As no hydrogen recombination lines are detected in
IRAS~22272+5435, no photons more energetic than Ly$\alpha$ or $h\nu$ =
10.2~eV are available. It is found that a grain should be smaller than
6~\AA~ in radius ($N_C < $ 100) to be heated to 760~K by a single
Ly$\alpha$ photon. We use the empirical specific heat of graphite
measured at a low temperature \citep{tou70} and the graphite mass
density 2200 kg m$^{-3}$ in estimating the heat capacity. This is very
small for a solid particle; however, grains of this size have been
considered as a common constituent of reflection nebulae to account
for the hot blackbody continuum of about 1000~K \citep{sel84}.

Alternatively, as a PPN is a mass-losing object, the observed spatial
variation could be interpreted as a record of the history of grain
production of the central star. The changing chemical composition at
the outer circumstellar shell could represent the temporal transition
of hydrocarbon output during the evolution of the PPN. The possible
mode-switching in the production of dust has been proposed by
\citet{bus93,geb97,kwo99}; and \citet{kwo01} to account for the
systematic difference of the IEF in PNs and their progenitor
PPNs. Thus, the thermal processing is not necessarily a current
ongoing event. Actually, the most obvious heat source in the nebula is
the central star and the warm circumstellar shell close to it. The
particular chemical composition, or a sign of thermal processing,
would be acquired when a grain was still in the warm environment near
the star.  We will call the inhomogeneous chemical composition
resulting from the temporal transition of grain formation the ``old
record'' scenario. In the scenario, the thermal processing was
completed close to the central star, and is being observed now when
the dust grains with varying chemical compositions are blown off to
1000~AU away from the star.

We will consider two heating scenarios in the following discussion,
(1) the ongoing stochastic heating of a very small grain, and (2) the
old record scenario in which the spatial variation in the chemical
compositions originates in the past at the time of grain formation in
the warm region near the central star.

\subsubsection{Preserving Aliphatic Material}

The immediate problem with the stochastic heating scenario is how the
aliphatic material could survive through the warmer environment closer
to the central star. The physical conditions are more favorable to
active thermal annealing there; for instance, at the temperature of
carbon condensation, the aliphatic material should be totally
graphitized.

Before discussing the preservation of the aliphatic material, first we
consider the location where the ultra-small grains could form. Because
of the lack of an efficient destructive mechanism, a grain
monotonically grows in the mass-loss wind as it travels through the
circumstellar shell \citep{gai88}. Thus, a small grain can be formed
only at the outer region of the envelope, where the grain growth is
very inefficient; otherwise, the final dimension would be too large
(Figure~\ref{prsv}).  \citet{dom89} found that while most of the
grains are nucleated at 1--2~$R_\ast$ from the star and keep growing
until they are as large as 0.01--0.1~$\mu$m, the population of small
grains of carbon content $N_C$ = 100 should not be nucleated at
7~$R_\ast$ or closer to the central star. The temperature there falls
down well below 600~K \citep{gai87}. It is cool enough to allow the
aliphatic C-H bonds to be preserved.

In the old record scenario, we do not have to assume the grains are
ultra-small. Most of the grains are formed in a thin shell near the
sonic point 1--2~$R_\ast$ from the star. The nucleation and growth
rates are very high there. However, grain formation is not an
instantaneous event. Although the growth rate drops by orders of
magnitude, a grain keeps growing even at tens of stellar radii from
the star until the source gas is completely depleted or diluted
\citep{dom89}. This means that the outermost layer of a grain particle
is formed in the cool region where thermal processing no longer
occurs. Thus, while the inner core formed in the warmer region is
already graphitized, the fresh surface of a grain is most likely
covered by the aliphatic C-H bonds (Figure~\ref{prsv}).  The growth of
an ``amorphous carbon mantle'' in the cool ($<$1100~K) mass-loss wind
has been discussed by \citet{gai84}.

\subsubsection{Mass-Loss History}

We have heat sources, and the raw material is ready for the thermal
treatment. Our observation shows less processed material inside, and
more processed outside. We discuss how the observed spatial variation
of hydrocarbon features could possibly be produced.

We start with a gradual annealing of a small grain when it travels
outward in the circumstellar shell. The incidence of the Ly$\alpha$
photon per grain from a G5Ia central star at the distance of 1000~AU
can be predicted using the stellar spectral model of \citet{kur79}.
Assuming all photons of $\lambda$ $\leq$ 1215\AA~are converted to
Ly$\alpha$ and the total luminosity of the central star to be
1.3$\times 10^4 L_\odot$ \citep{uet01}, the photon rate to a 6~\AA~
particle turns out to be about once in thousands of years.  However,
it critically depends on the wavelength range we consider. For
instance, if we count the photons $\lambda$ $\le$ 1500~\AA, the
incidence rate jumps up to once in every 5 yr. It drops by an order or
two when visible extinction of A$_V$ = 1 is applied. At each
incidence, a grain cools down in approximately 10$^{-1}$ s while
emitting as a blackbody. As a grain is blown off by roughly 1~AU per
year if a 10~km~s$^{-1}$ mass-loss velocity is assumed; the region of
our interest, 600--1300~AU from the star, corresponds to $\sim$700~yr
for a grain to cross. The Ly$\alpha$ photon rate ($\lambda \leq$
1215\AA) without extinction does not conflict with the grain traveling
time. The higher fraction of the dust particles have experienced more
energetic photon incidence in the outer region than in the inner
region because they have been soaked in the radiation field longer.
Hence, we would see more thermally annealed grains closer to the edge
of the nebula. However, we have to be cautious about the coincidence
of the timescale. First, we implicitly assumed the effective
temperature of the central star does not significantly change during
the relevant period of the time, which is not a trivial issue. Second,
the reorganization of the carbon frameworks may not finish in an
instant, but may take a finite period of time until the aliphatic to
aromatic transformation is settled down.  When the alteration
timescale is longer than the cooling time, a single Ly$\alpha$ photon
is insufficient to transform the chemical composition of the entire
grain. Further discussion on the gradual annealing scenario would be
premature until the microscopic processes of thermal annealing, as
well as the effect of the evolution of the central star, is fully
understood.

A size variation with distance from the star may provide a solution
that accounts for the spatial variation in the chemical
composition. Suppose that there is a dominant population of small
grains farther away from the star; they are more processed than inner
ones because a smaller grain reaches higher temperatures when heated
with same energy input. Such a size variation with the distance from
the star have been often observed in other PPN (e.g., Sahai et
al. 1998). As the higher mass loss rate results in the larger grain
size \citep{kru97}, the accelerating mass loss at the end of AGB might
be relevant to the production of the hypothetical grain size
distribution.

On the other hand, in the old record scenario the assumed grain size
is large enough to keep grains from thermal fluctuation. Thus, once
the grain growth ceases near the star ($<$ 20~$R_\ast$), the chemical
composition of the grain surface remains unchanged during the traverse
in the mass-loss wind. Then the variation of the spectral feature we
observed at 1000~AU from the star simply reflects how the grain
surface was composed near the star. If the timescale of grain growth
is slow, it may keep growing in the cool region in the outer
circumstellar shell, gradually building more aliphatic-rich mantles on
the surface. By contrast, if the source gas quickly depletes closer to
the star, a grain stops growing earlier in the warm region, and the
surface composition remains mainly graphitic. The observed spatial
variation implies a history of the gradual switching of these dust
formation modes. However, dust formation models available at present
are based on the stationary mass-loss wind, which is ineffective for
predicting the temporal variation of chemical composition in dust
production. The mass-loss rate is very sensitive to the stellar
parameter at the end of the AGB phase (e.g., Willson 2000). In
particular, in the superwind phase the mass loss shows complicated
temporal variation strongly dependent on the initial mass of stars
\citep{wac02,schro99}. To fully understand how time-variable mass loss
affects the chemical properties of dust, how those dust grains build
up a heterogeneous spatial structure around the central star, and how
eventually the spectrum of the chemical composition of dust becomes
the carbon-rich late-type stars return to the ISM, further theoretical
and observational studies are needed.

\section{Summary}
We presented spatially resolved 3~$\mu$m spectroscopy of the
proto-planetary nebula IRAS~22272+5435 that reveals the different
distribution of the hydrocarbon gas and dust. The main conclusions are
follows:

1. The acetylene (C$_2$H$_2$) and hydrogen cyanide (HCN) absorption is
   found in the central star spectrum at 3.0~$\mu$m. The molecules are
   highly concentrated in the compact region at the center.

2. The 3.3 and 3.4~$\mu$m hydrocarbon dust emission is detected at
   600--1300~AU from the star. The spatial separation of C$_2$H$_2$
   and hydrocarbon dust is observed for the first time, and it
   reinforces the classical view of the dust formation.


3. We found the relative intensity of the aliphatic feature at
   3.4~$\mu$m to the aromatic feature at 3.3~$\mu$m decreases with the
   distance from the star. The spatial variation of the spectral
   feature is well reproduced in the spectra of a laboratory analog of
   carbon dust. The thermal process is likely to account for the
   spectral variation.
   
4. We suggested the stochastic heating of very small grains ($a < $
   6~\AA) or the grain surface composition at the end of grain growth
   would explain the changing aliphatic and aromatic spectral
   features. In both cases the time-dependent mass-loss profile should
   be the key to reproducing the spatial variation.
   
\acknowledgments 

We thank all the staff and crew of the Subaru Telescope and NAOJ for
their valuable assistance in obtaining these data and continuous
support for IRCS and Subaru AO construction. Takashi Kozasa and Masao
Saito are appreciated for their useful suggestions in the
discussion. We are grateful to V. Mennella and collaborators for
making their data available for use in Figure \ref{men}. We are also
grateful to J. E. Chiar and collaborators for kindly allowing us to
reproduce their data in Figures \ref{mol2} and \ref{men}. Special
thanks goes to A. T. Tokunaga for many inspiring discussions and
enduring encouragement during the writing. We thank the anonymous
referee for the helpful comments that make the manuscript more
consistent and readable. M. Goto is supported by a Japan Society for
the Promotion of Science fellowship. Last, but not least, we wish to
express our appreciation to those of Hawaiian ancestry on whose sacred
mountain we are privileged to be guests.

\clearpage

\figurenum{1}
\begin{figure}
\caption{Top: A blow up of a pair-subtracted spectrogram of
IRAS~22272+5435 near the 3.4~$\mu$m region. The spectrograms were
recorded at two different locations of the array (position A and
position B) by nodding the AO tip-tilt mirror by 2\arcsec. The two
spectrograms are subtracted from each other to remove the continuum
and the line emissions of the sky. Bottom: A cross-cut of the positive
and negative spectrograms at 3.4~$\mu$m. The slit was put along the
position angle 56$^\circ$ along the elongation of the mid-infrared
image. The apertures of 4 pixels (0\farcs23) in width are defined 
every 2 pixels. The size and location of apertures are shown by
boxes. \label{prof}}
\end{figure}

\figurenum{2}
\begin{figure}
\caption{The spectra of IRAS~22272+5435 extracted from the apertures
shown in Figure~\ref{prof}. Left: From the spectrogram obtained at the
position A. Right: From position B. Power-law continua are defined by
fitting the wavelength regions free of spectral features. All spectra
are normalized at 3.5~$\mu$m. The off-center spectra are offset
relative to the central star spectrum for clarity, and placed at every
0.5 grid in the relative intensity unit. \label{spec1}}
\end{figure}

\figurenum{3}
\begin{figure}
\caption{Same as Figure~\ref{spec1}, but binned by 4 pixels, and the
power-law continua are subtracted. While the spectra close to the
central star are free from dust emission features, the molecular
absorption features of C$_2$H$_2$ and HCN are found (See
Fig. \ref{mol2}). The broad emission features centered at 3.3~$\mu$m
and 3.4~$\mu$m are of aromatic and aliphatic hydrocarbons,
respectively. The off-center spectra are offset relative to the
central star spectrum and placed in every 0.25 grid in the relative
intensity unit. \label{spec2}}
\end{figure}

\figurenum{4}
\begin{figure}
\caption{A comparison of the 3.0~$\mu$m absorption feature of
IRAS~22272+5435 at the central star with the acetylene (C$_2$H$_2$)
and hydrogen cyanide (HCN) bands found in CRL~618 \citep{chi98}. The
C$_2$H$_2$ and HCN spectra are also calculated using the HITRAN
database \citep{rot98} and shown in the bottom of the
panel. \label{mol2}}
\end{figure}

\figurenum{5}
\begin{figure}
\caption{Left: A blow-up of the hydrocarbon emission features in the
northeast region of the nebula. The spectra are continuum subtracted
and normalized at 3.3~$\mu$m. Right: A sequence of absorption spectra
of laboratory analogs of hydrocarbon dust subject to thermal annealing
\citep{got00}.\label{labo}}
\end{figure}

\figurenum{6}
\begin{figure}
\caption{A comparison of the 3.4~$\mu$m {\it absorption} features
toward CRL~618 \citep{chi98} and Galactic center source GCS~3
\citep{chi00} with the 3.4~$\mu$m {\it emission} feature of
IRAS~22272+5435. Two types of laboratory analogs from \citet{men99}
and \citet{got00} are shown for comparison. The vertical dotted lines
indicate the asymmetric vibrational modes of C-H bonds in methyl
(3.38~$\mu$m) and methylene groups (3.42~$\mu$m), and the symmetric
vibrational mode of methylene groups (3.48~$\mu$m)
\citep{dis83}. \label{men}}
\end{figure}

\figurenum{7}
\begin{figure}
\caption{A schematic of how the aliphatic material could be formed and
preserved in the warm circumstellar environment. A small grain starts
becoming nucleated only at a distance from the star where the ambient
temperature is cool enough not to destroy the aliphatic C-H bonds. The
most of the grains originate in the area warm and close to the central
star, and rapidly grow as they are blown away with the mass-loss
wind. Because the grain growth continues until the source gas is
completely depleted, the aliphatic mantles gradually form around their
graphitized cores. \label{prsv}}
\end{figure}

\clearpage

\figurenum{1}
\begin{figure}
\plotone{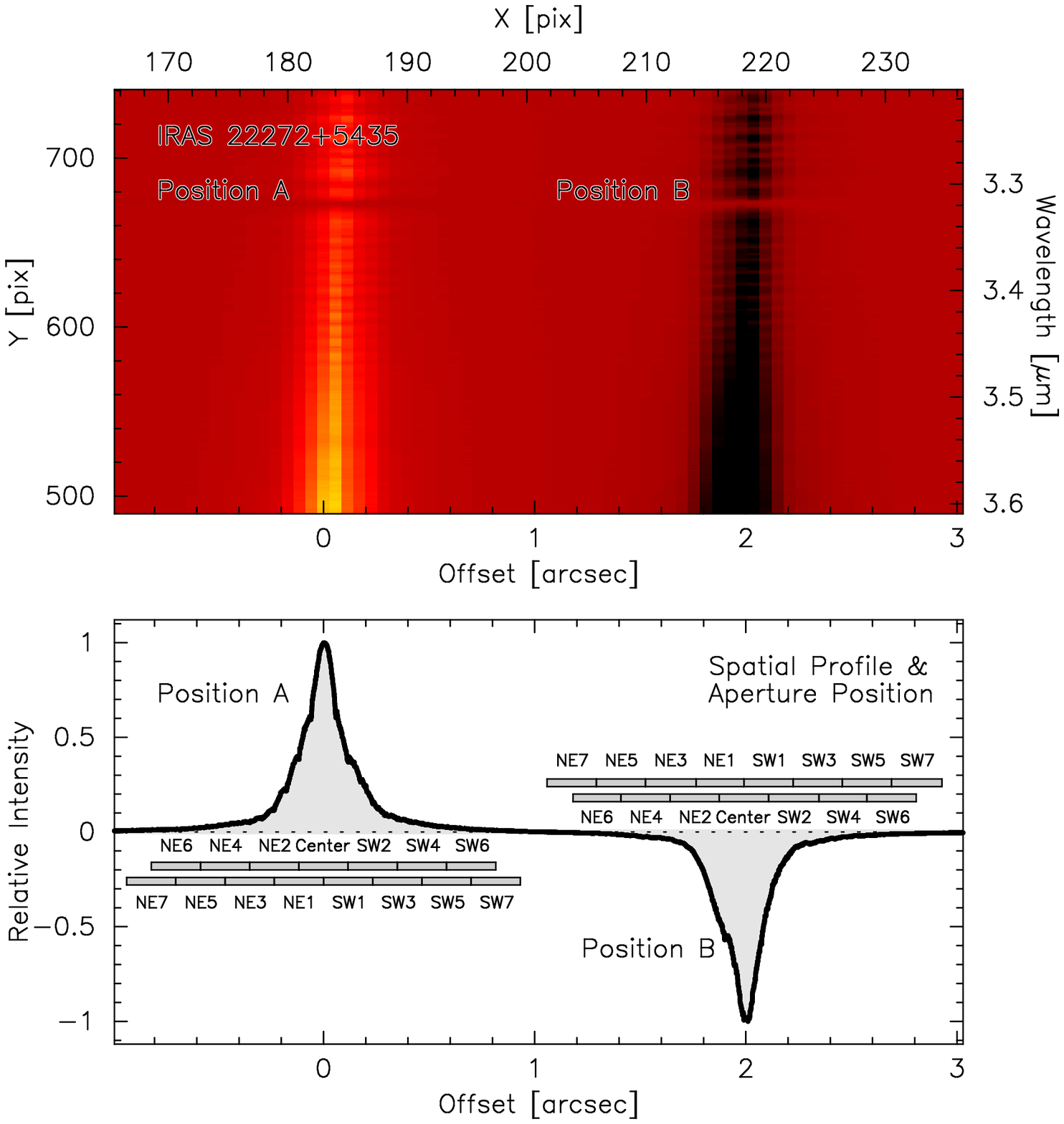}
\caption{}
\end{figure}

\figurenum{2}
\begin{figure}
\plotone{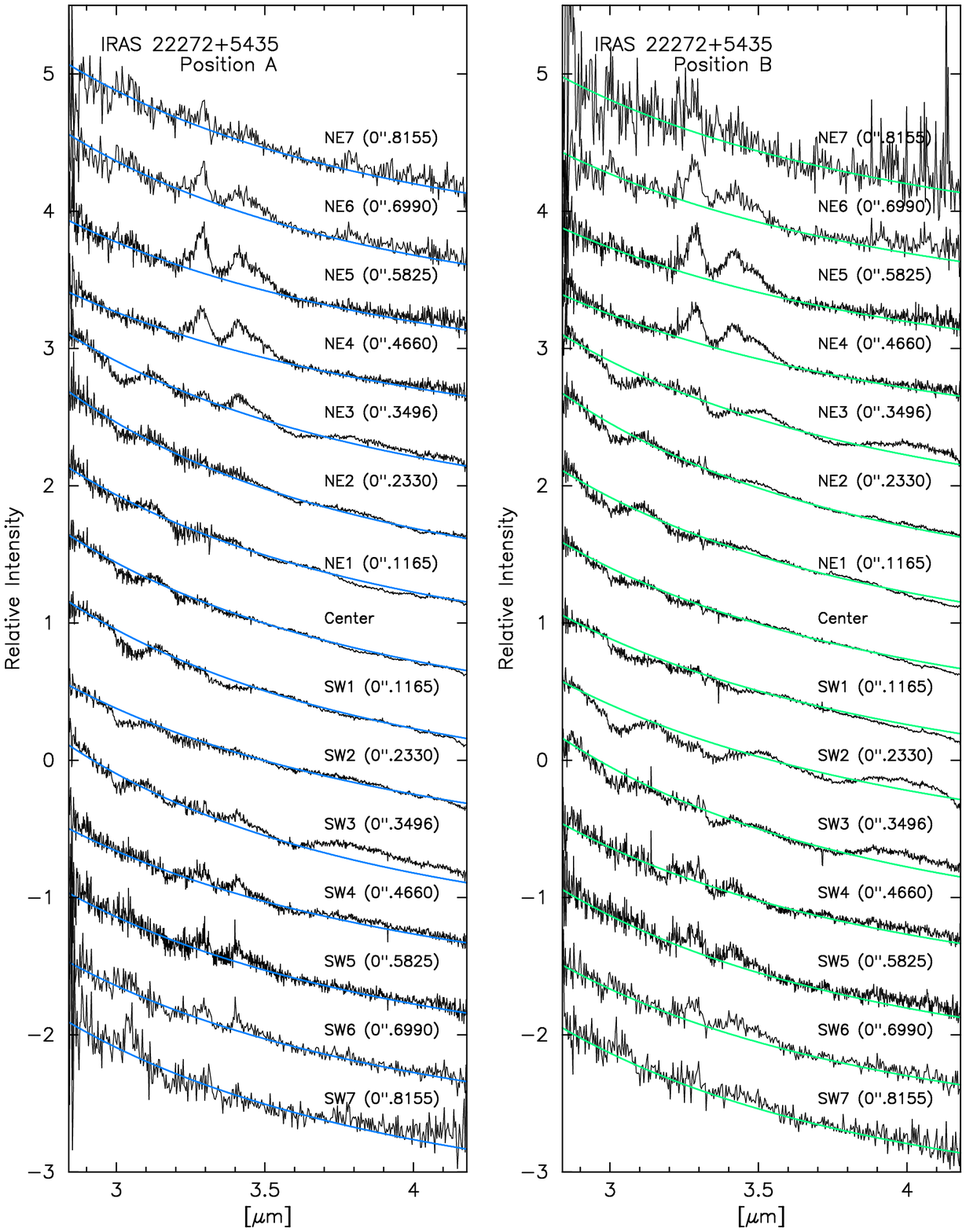}
\caption{}
\end{figure}

\figurenum{3}
\begin{figure}
\plotone{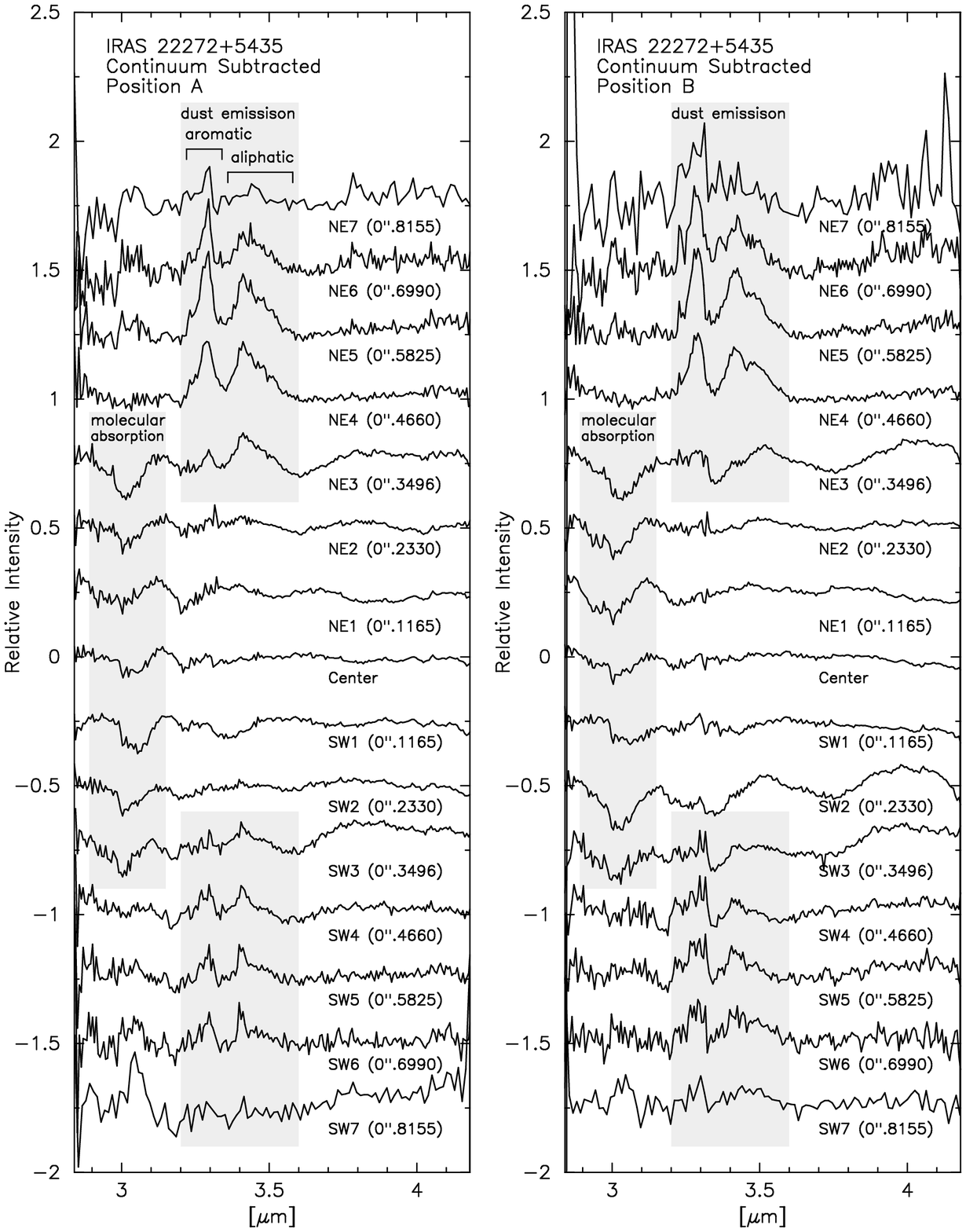}
\caption{}
\end{figure}

\figurenum{4}
\begin{figure}
\plotone{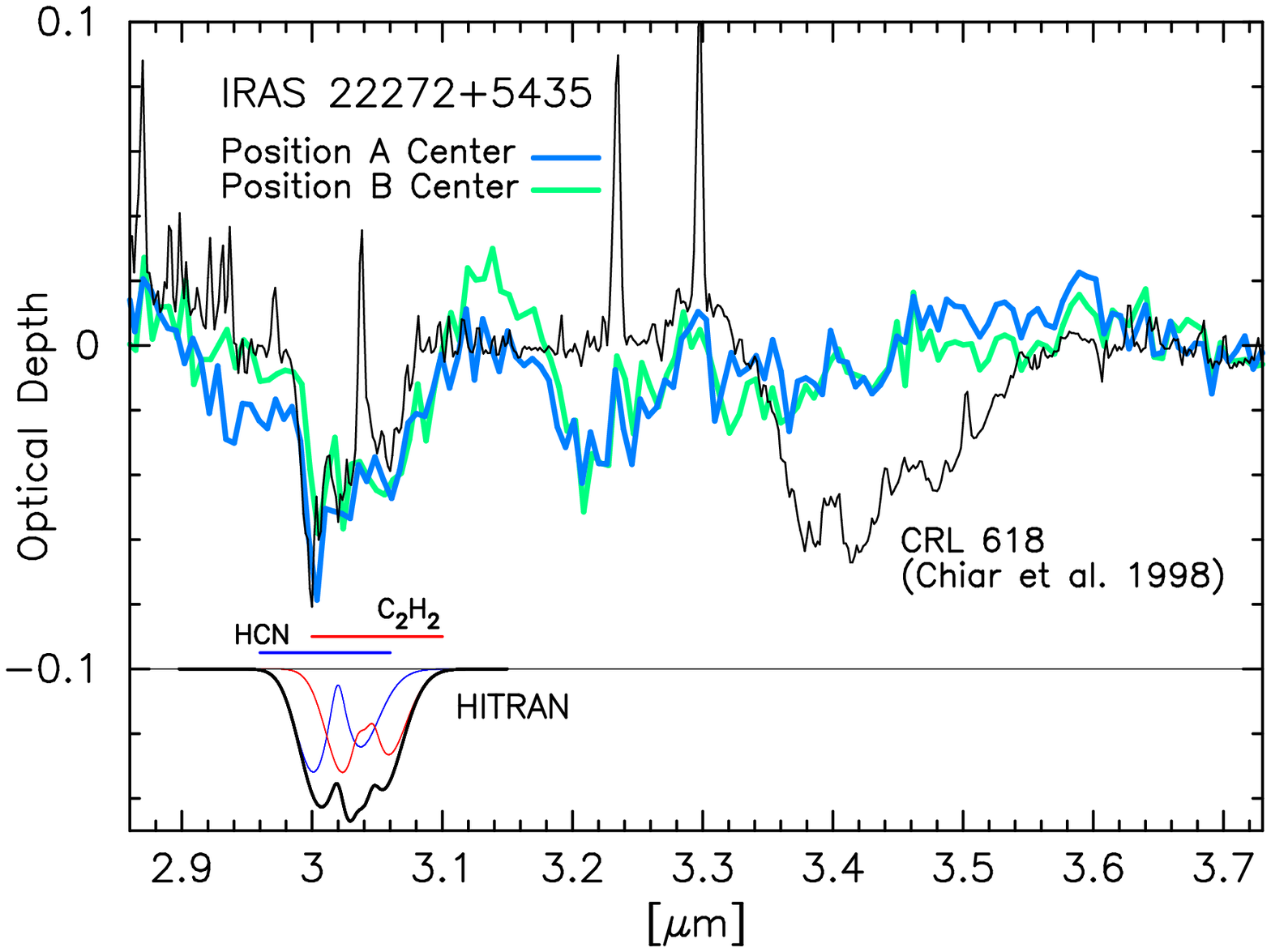}
\caption{}
\end{figure}

\figurenum{5}
\begin{figure}
\plotone{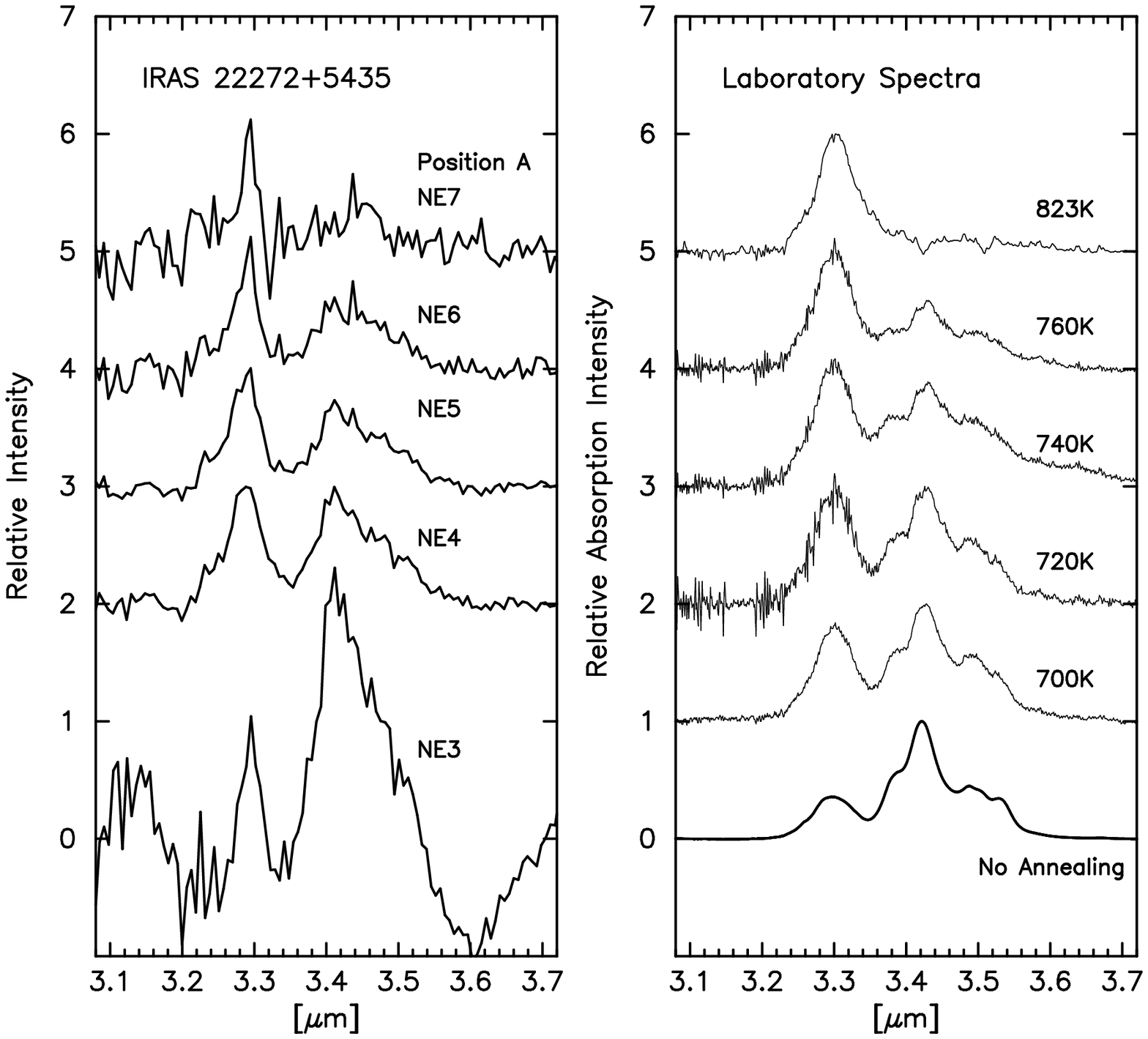}
\caption{}
\end{figure}

\figurenum{6}
\begin{figure}
\plotone{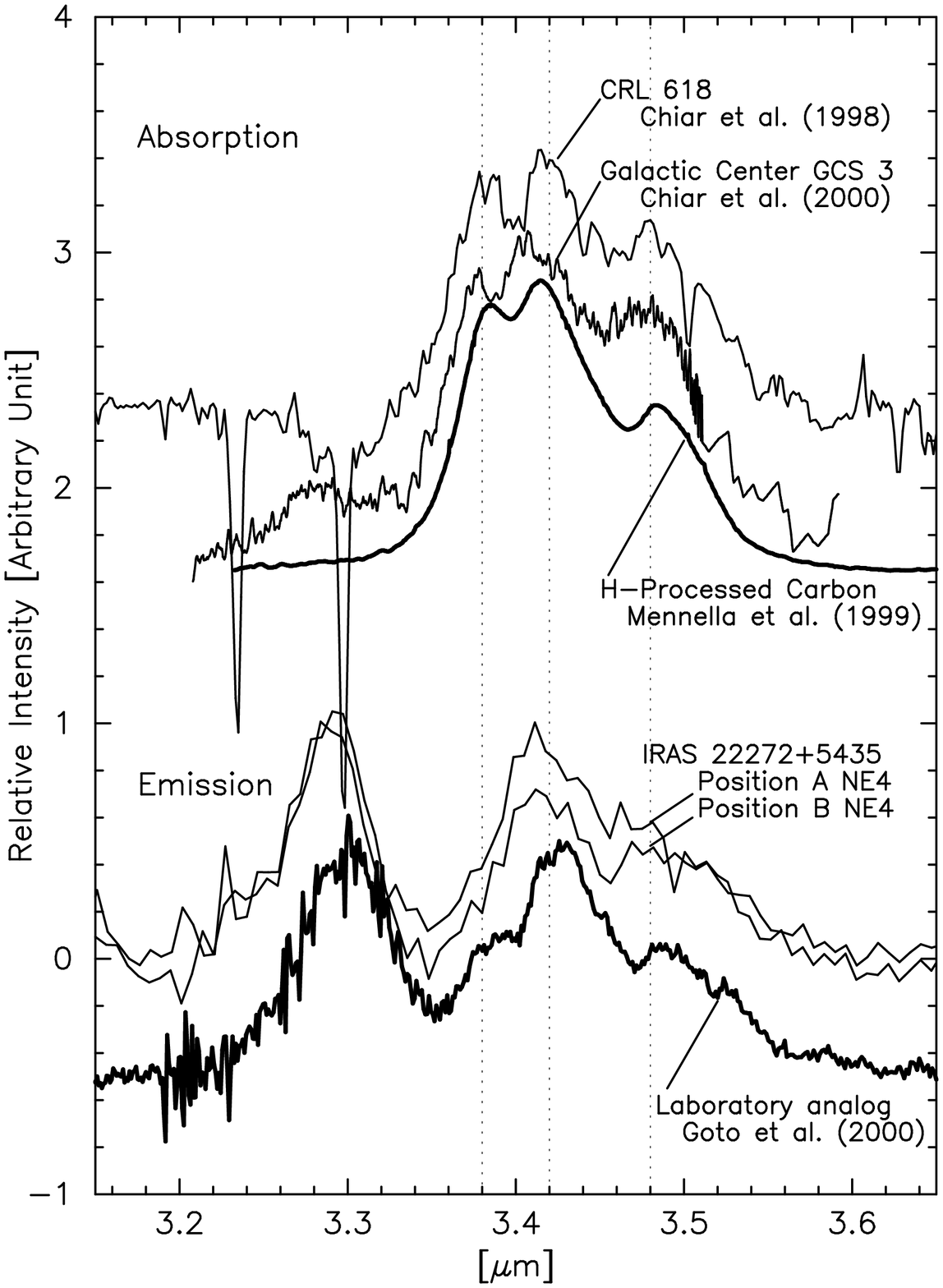}
\caption{}
\end{figure}

\figurenum{7}
\begin{figure}
\plotone{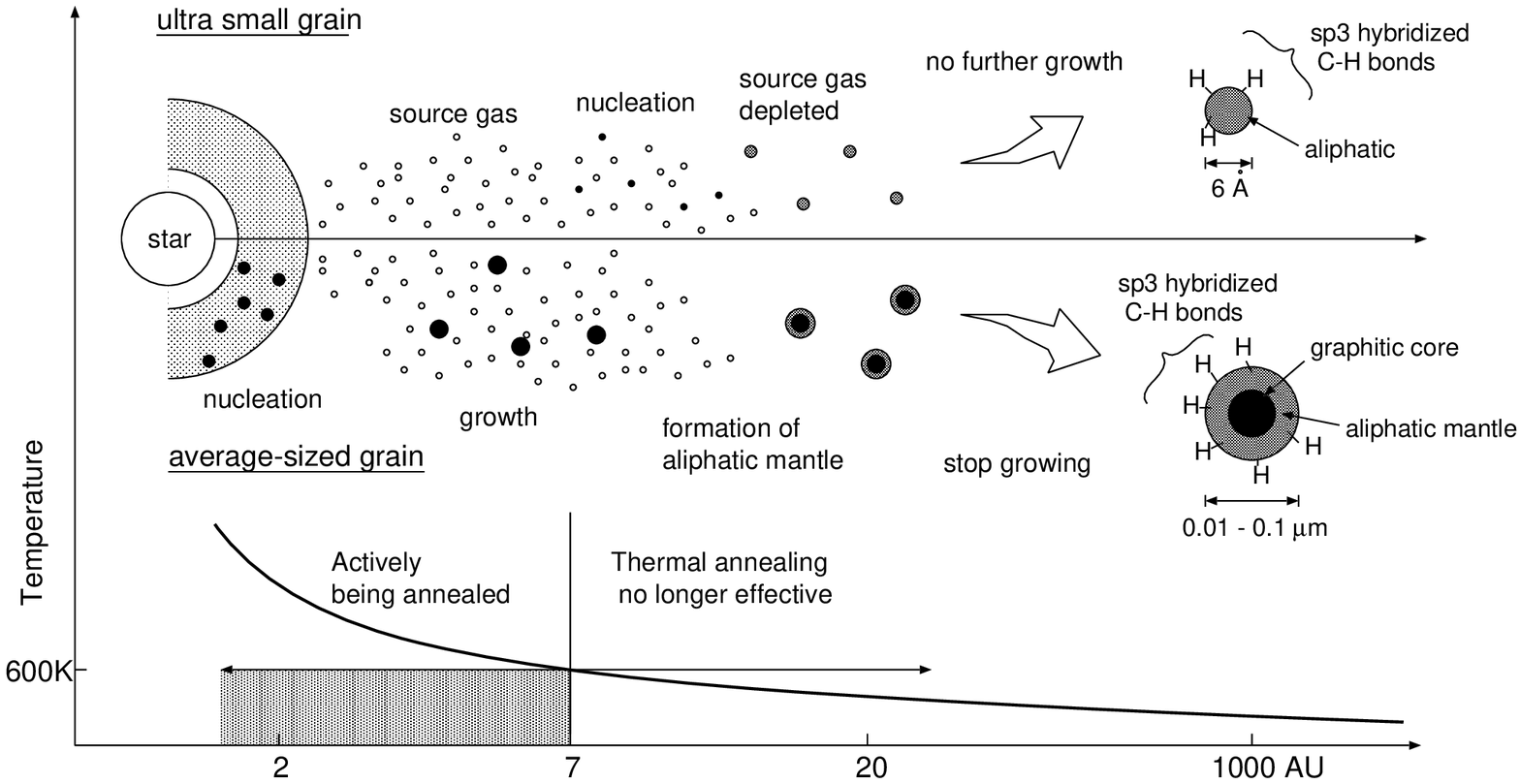}
\caption{}
\end{figure}

\end{document}